\documentclass[twocolumn,showpacs,preprintnumbers,amsmath,amssymb]{revtex4}
\newcommand{\BEQ}{\begin{equation}}
\newcommand{\EEQ}{\end{equation}}
\newcommand{\BEA}{\begin{eqnarray}}
\newcommand{\EEA}{\end{eqnarray}}

\renewcommand{\d}{{\rm d}}

\newcommand{\Sig}{\Sigma}

\newcommand{\half}{\frac{1}{2}}

\newcommand{\QQ}{{\cal Q}}

\renewcommand{\thesection}{\arabic{section}}

\def\dbarrm {{\mathchar'26\mkern-11mu{\rm d}}}                       %
\begin{document} 
\draft
\title
{On testing the violation of the Clausius inequality 
in nanoscale electric circuits}
\author{A.E. Allahverdyan$^{1,2)}$ and Th.M. Nieuwenhuizen$^{2)}$}
\address{
$^{1)}$Yerevan Physics Institute,
Alikhanian Brothers St. 2, Yerevan 375036, Armenia;\\ 
$^{2)}$ Institute for Theoretical Physics,
University of Amsterdam,
Valckenierstraat 65, 1018 XE Amsterdam, The Netherlands}

\begin{abstract}
The Clausius inequality, one of the classical formulations of the second 
law, was recently found to be violated in the quantum regime.
Here this result is formulated in the context of a mesoscopic or nanoscale
linear RLC circuit interacting with a thermal bath. 
Previous experiments in this and related fields are analyzed and 
possibilities of experimental detection of the violation are pointed out.
It is discussed that recent experiments reached the range of 
temperatures, where the effect should be visible, and that
a part of the proposal was already confirmed.
\end{abstract}
\pacs{03.65.Ta, 03.65.Yz, 05.30}

\maketitle


\renewcommand{\thesection}{\arabic{section}}
\section{ Introduction}
\setcounter{equation}{0}\setcounter{figure}{0} 
\renewcommand{\thesection}{\arabic{section}.}

The application of thermodynamics to electric circuits
has a long and remarkably fruitful history \cite{ny,leon,landauer,
meixner,strat,chu}. In the late twenties, by applying general principles
of thermodynamics | in particular, the second law | Nyquist \cite{ny} 
deduced the spectrum of the fluctuation force acting in an equilibrium 
electric circuit. This result was much later confirmed by microscopic 
approaches \cite{strat,haake,denker, weiss,klim} and became known as 
the Nyquist spectrum. 
Nearly twenty years later Brillouin \cite{leon} applied the second law to
analyze a circuit containing rectifying elements. Another formulation
of the second law, the Clausius inequality, was considered in the context 
of electric circuits by Landauer \cite{landauer}. 
The equilibrium thermodynamics
of linear and non-linear circuits was thoroughly analyzed by Stratonovich
\cite{strat}. Further research in this field was stimulated by
two facts: first, by the technical importance of circuits in electronics,
and second by their feasibility, which allows to create experimental 
conditions close to those in theory \cite{chu}. 

In view of these successful applications of thermodynamics, one naturally
expects that electrical circuits can play also a complementary role 
by acting as experimental and theoretical laboratories for testing new 
ideas and results in statistical thermodynamics itself. 
The present paper makes such an attempt in the context of our recent 
discussion of the applicability of the second law to quantum systems 
coupled to thermal baths \cite{AN,kapur}. 
The general philosophy of the approach is that thermodynamic relations are 
not introduced axiomatically or phenomenologically, but should be derived 
from first principles, namely the laws of quantum mechanics. 
For linear systems, e.g. a set of harmonically bounded Brownian particle 
interacting with a quantum thermal bath, this program can be carried out 
exactly.
As the main result we were able to check some formulations of the second law,
whose validity in the classical domain was numerously confirmed via 
analogous approaches \cite{meixner,landauer,strat}. 
One of these formulations, the Clausius inequality, appeared to be broken 
in the low temperature quantum regime.
Here we reformulate this result for a quantum linear
RLC circuit \cite{haake, denker,serge}. 
Our purpose is rather straightforward: we explain that 
the above violation can be detected experimentally in low temperature 
mesoscopic circuits. To this end we analyze some known experimental results
and show that several important parts of our proposal were already realized
in experiment. 

Our plan for the present paper is following. 
In section 2 we will briefly describe the 
quantum RLC circuit coupled with a thermal bath.
We continue with explanation
of the Clausius inequality in section 3. In the following section we
analyze some experimental results, and in their context we make 
quantitative estimates for our effect. Our conclusions are presented 
in the last section.

\renewcommand{\thesection}{\arabic{section}}
\section{ RLC circuit and its Heisenberg-Langevin equations}
\setcounter{equation}{0}\setcounter{figure}{0} 
\renewcommand{\thesection}{\arabic{section}.}

\subsection{Classical RLC circuit}

The classical scheme of the simplest RLC circuit is well-known.
It consist of capacity $C$, inductivity $L$ and resistance $R$.
The loss of voltage across the resistance is given by the Ohm law: $IR$,
where $I=\d Q/\d t$
is the current, and $Q$ stands for the charge. The capacitor enters 
the total voltage as $Q/C$. Finally, the inductive element induces a
magnetic field with the flux $\Phi = L I$, which in turn contributes to 
the voltage (Faraday's law). Altogether, one finally obtains:
\BEA
\dot{Q}=\frac{\Phi}{L},\qquad
\dot{\Phi}=-\frac{Q}{C}-\frac{R}{L}\,\Phi.
\label{2}
\EEA
The first equation is just the definition of the current, and the second
one expresses the fact that the total voltage in the closed circuit is zero.
Apart from the term connected with the resistance, Eqs.~(\ref{2})
can be viewed as the canonical equation of motion generated by the Hamiltonian
\BEA
\label{s}
H_{S}=\frac{\Phi^2}{2L}+\frac{Q^2}{2C},
\EEA
where $Q$ and $\Phi$ are the canonical coordinate and momentum. 
Within the language of Brownian motion the 
contribution of the resistance in Eq.~(\ref{2}) corresponds to the Ohmic 
friction with the damping coefficient $R$. 
In the same context $C$ corresponds to the inverse strength of the 
external harmonic potential, and $L$ corresponds to the mass of the 
Brownian particle.

\subsection{Quantum RLC circuit}

Eq.~(\ref{s}) makes obvious that for $R=0$ one can quantize 
the model regarding $\Phi$ and $Q$ as the corresponding operators:
\BEA \label{QPcom}
[Q, \Phi]=i\hbar.
\EEA
Then Eqs.~(\ref{2}) are just the Heisenberg equations of the problem. 
The quantum description for electrical circuits
became necessary at the beginning of 1980's with the 
appearance of gravitational wave-measuring setups
and Josephson junctions. 
These devices operate at low temperatures and are very susceptible to their
environment, so that both the circuit and its photonic
thermal bath have to be described quantum-mechanically.
Since then the problem of quantization for the electrical circuits was 
considered in numerous contributions
(see e.g. \cite{haake,denker,serge}) with  special emphasis on the dissipative
aspects of the problem. More recently the interest in this subject was
renewed in the context of low-temperature mesoscopic circuits 
~\cite{w1,flabby}.
Though within the classical approach the resistivity can be introduced 
phenomenologically, this is impossible for the quantum case, in particular
because it will violate the Heisenberg relation.
The cause is that even if Eq. (\ref{QPcom}) is valid at the initial moment, 
a non-Hamiltonian dynamics does not conserve it in time. Thus, the
dissipative quantum situation should be investigated starting from a more 
fundamental level, i.e. by explicitly describing  the thermal bath. 
The strategy here 
is exactly the same as when studying the dynamics of open quantum systems
in general \cite{weiss}: One models the resistance as an open chain of linear
LC circuits (thermal bath) attached to the studied circuit, and then applies 
the standard canonical quantization scheme to the whole closed Hamiltonian
system. In a second step one traces out the bath, since only the degrees of 
freedom of the initial
circuit are considered to be observable. Since the bath consists of harmonic 
oscillators, this procedure can be realized explicitly.
Omitting technicalities which can be found in \cite{haake,denker,weiss,AN}, 
we will write down the final quantum Langevin equations
\BEA
&&\dot{Q}=\frac{\Phi}{L},\\
&&\dot{\Phi}=-\frac{Q}{C}
-R\,\Gamma\int _{0}^{t}\d s\, e^{-\Gamma(t-s)}
\dot{Q}(s)+\eta (t)\nonumber\\
&&-R\,\Gamma\,e^{-\Gamma t}\,Q(0),
\label{4}
\EEA
where $\Gamma$ is the maximal frequency of the bath, and 
where $\eta (t)$ is the quantum Gaussian noise (random e.m.f.) 
with the Nyquist spectrum:
\BEA
\label{ppp} 
&&K(t-t')=\frac{1}{2}\langle \eta (t) \eta (t') +  
\eta (t') \eta (t) \rangle \\
&&=\frac{\hbar R }{\pi}\int_0^{\infty}\d \omega 
\,\frac{\omega \, \coth \left (\half\beta\hbar\omega\right)}
{1+(\omega /\Gamma )^2}\,
\cos\omega (t-t'). \nonumber
\EEA
where $\beta=1/k_BT$, and we use units in which Boltzmann's constant
$k_B=1$. 

If $\Gamma$ is much larger than other frequencies of the problem
(this is the most typical situation), then for $t>0$ one can 
get the Langevin equation (\ref{4}) in a more standard form:
\BEA
\label{maximus}
\dot{\Phi}=-\frac{Q(t)}{C}-R\,I(t)+\eta (t).
\EEA
In the classical limit (large $T$) the spectrum (\ref{ppp}) would become
the Nyquist white noise spectrum
\BEA 
K(t-t')=RT\,\Gamma e^{-\Gamma|t-t'|}\approx 2RT\delta(t-t'),
\EEA
but that regime will not be of our concern.
Though in the classical situation the noise can be omitted at $T=0$,
for the quantum case the presence of a resistivity without 
the corresponding noise is excluded. 

Generally, one should keep the parameter $\Gamma$ in Eq.~(\ref{ppp}) for the
noise correlation function, since otherwise some divergences will occur. 
However, provided that $\Gamma$ is large the concrete form of the cutoff 
function (here taken to be Lorentzian) is not essential \cite{AN,weiss}.

\subsection{Stationary state of the circuit}

Eq.~(\ref{maximus})
is linear, and can be solved exactly. We will not repeat the 
derivation of this solution, since it was thoroughly investigated in
\cite{AN,weiss}. Starting from any initial state the circuits relaxes to 
its stationary state, where $\Phi $ and $Q$ are independent 
random Gaussian quantities with zero averages:
$\langle \Phi\rangle=\langle Q\rangle =0$, and 
have the following dispersions ~\cite{haake}
\BEA
&&\langle \Phi^2\rangle =\int \frac{\d \omega}{2\pi }\,
\frac{\omega ^2\,k(\omega )}
{(1+\omega ^2/\Gamma^2)[\,(\omega ^2-\omega_0^2)\,^2+\omega^2R^2/L^2]},
\nonumber\\
&&\label{gri1} \langle Q^2\rangle = \int
\frac{\d \omega}{2\pi }\,
\frac{k(\omega )}{(\omega ^2-\omega_0^2)^2L^2+\omega^2R^2},\\
&&k(\omega ) = \hbar R  ~\omega ~{\rm coth}\frac{\hbar \omega }{2T},
\nonumber\EEA
where $\omega_0=1/\sqrt{LC}$ is the  frequency of the free circuit.
Statistically these variables are independent, which is expressed 
by the relation $\langle Q\Phi+\Phi Q\rangle=0$. 
Explicit formulas expressing $\langle Q^2\rangle$ and
$\langle \Phi^2\rangle$ in terms of di-gamma functions, 
are given in \cite{AN,kapur}. 

The disorder present in the circuit is characterized by the 
occupied phase-space volume 
\BEA
\Sig=\frac{\Delta \Phi\Delta Q}{\hbar}\equiv
\sqrt{\frac{\langle \Phi^2\rangle\langle Q^2\rangle}{\hbar^2} }.
\label{sig}
\EEA
The lower bound  $\Sig=\half$ follows from the Heisenberg relation
$\Delta \Phi\Delta Q\ge \half\hbar$. 
It means that the charge and the flux fluctuate close to their 
average values. 

It is important to notice that in general
the dispersions are not equal to their Gibbsian values: 
\BEA
\label{gi}
&&\langle \Phi^2\rangle_{\rm G}=
\half L\,\hbar\omega_0\tanh\half\beta\hbar\omega_0,\nonumber\\
&&\langle Q^2\rangle_{\rm G}=\half C\,\hbar\omega_0
\tanh\half\beta\hbar\omega_0,
\EEA 
which are obtained by assuming a Gibbs distribution for the circuit, 
valid for a weak coupling with the bath, i.e. when taking $R\to 0$. 
That is why $\langle \Phi^2\rangle_{\rm G}$ and
$\langle Q^2\rangle_{\rm G}$ do not contain the resistance $R$ anymore,
in contrast to the general expressions for
 $\langle \Phi^2\rangle$ and $\langle Q^2\rangle$, presented above. 

It is natural to identify the average energy stored in the circuit with
\BEA
\label{u}
U\equiv \langle H_S\rangle=
\frac{\langle \Phi^2\rangle}{2L}+\frac{\langle Q^2\rangle}{2C}.
\EEA

There is a general argument why the dispersions
$\langle \Phi^2\rangle$ and $\langle Q^2\rangle$ 
are not equal to their Gibbsian values \cite{AN,kapur}. 
For $T\to 0$ the Gibbs distribution predicts that the
circuit is in the ground state of its Hamiltonian $H_S$. Indeed, it can be
checked that when values (\ref{gi}) are inserted into (\ref{u}), one gets
$U=\half\hbar \omega_0$, just the exact ground state energy of 
the free (i.e. $R=0$) circuit. 
In quantum mechanics two interacting system are typically not in pure states,
even though the overall state of the total system may be pure.
This is the intriguing property of quantum entanglement. Thus, we should 
not expect that a quantum circuit interacting non-weakly 
with its low temperature bath will be found in a pure state.
The approximate equalities 
$\langle \Phi^2\rangle\approx\langle \Phi^2\rangle_{\rm G}$, 
$\langle Q^2\rangle\approx\langle Q^2\rangle_{\rm G}$
are valid only for two particular cases:
the weak-coupling situation, where in (\ref{gri1}) one takes $R\to 0$,
and the classical case $\hbar/T\to 0$, where the temperature of the bath is so 
high that all signs of the quantum effects disappear. 
In both these situations the entanglement is very weak.

\renewcommand{\thesection}{\arabic{section}}
\section{Clausius inequality}
\setcounter{equation}{0}\setcounter{figure}{0} 
\renewcommand{\thesection}{\arabic{section}.}

Let one of the parameters of the circuit (e.g. the inductivity $L$) 
be varied  by an external source from $L$ to 
$L+\d L$, in a certain time interval. 
The variation is assumed to be very slow, so
that at any moment the distributions of the flux and the charge are 
still given by (\ref{gri1}) with the instantaneous inductance $L=L(t)$.
The variation itself is a accompanied by the work done by the external source.
A part of that work is stored in the circuit, and the rest is
transferred to the bath as heat. 
The energy budget of the variation is given by the first law:
\BEA
\label{carbon}
\frac{\d U}{\d L}=\frac{\dbarrm W}{\d L}+
\frac{\dbarrm \QQ}{\d L},\qquad \frac{\dbarrm W}{\d L}=
\left\langle \frac{\partial H_S}{\partial L}\right\rangle
=-\frac{\langle \Phi^2\rangle}{2L^2},
\EEA
where $\d U$ is the change of the energy stored in the circuit, $\dbarrm W$
is the work done by external source on the system, and the difference 
between them, the heat $\dbarrm \QQ$, is the energy
that goes from the bath to the system \cite{keizer,klim,balian,AN}.

Thermodynamics imposes a general relation between the heat received 
by the circuit and the change of its phase-space volume $\d \Sig$.
This statement was proposed by Clausius in the last part of the
nineteenth century, and  became established as one
of the formulations of the second law \cite{klim,balian}. 
There are several levels of mathematical rigor by which the
Clausius formulation can be presented \cite{keizer,klim,balian,AN}. 
For our present purposes it will be enough to use 
the simplest version \cite{AN,kapur}: If the the circuit receives from 
the bath a positive amount of heat $\dbarrm \QQ>0$, then its phase-space volume
is increased: $\dbarrm \Sig>0$. On the other hand, if the circuit is subjected
to the squeezing of its phase-space volume: $\d \Sig <0$, then it has to
release heat to the bath: $\dbarrm Q<0$. In formulas it reads:
\BEA
\label{c}
&&\qquad \dbarrm \QQ>0\qquad \Rightarrow\qquad \d \Sig >0;\nonumber\\
&&\qquad \d\Sig <0 \qquad \Rightarrow\qquad \dbarrm \QQ<0.
\EEA
In the classical domain everybody had a chance to observe the validity 
of the Clausius formulation
when looking at a squeezed substance which heats its environment (e.g. a
working pomp), or at a heated substance which tends to increase its volume 
(e.g. boiling water).
For the reader who is familiar with the formal structure of thermodynamics 
we mention that the Clausius formulation can be presented as 
the Clausius inequality
$\dbarrm \QQ\le T\,\d S$, where $S$ is the entropy of the system 
For our circuit the so-called von Neumann entropy reads~\cite{AN}
\BEA\label{SvN=}
S =(\Sigma+\half)\ln(\Sigma+\half) -(\Sigma-\half)\ln(\Sigma-\half),
\EEA
which is well a behaved function, since, as we discussed, 
the variable $\Sigma$ is larger than or equal to $\half$. It starts at
$S(\half)=0$, increases monotonically, and behaves for large $\Sigma$ as 
$S=\ln\Sigma+1+{\cal O}(1/\Sigma)$.

Eqs.~(\ref{c}) follow from assuming $\dbarrm Q\le T\d S$ upon noticing 
$\d S\propto +\,\d \Sig$ \cite{AN,kapur}. 
In particular, for $T=0$ this inequality
produces another version of the Clausius formulation: No heat can be
extracted from a zero temperature thermal bath.
The remaining inequality $\QQ(T=0)\le 0$ says that heat can only be 
dumped into the bath.

As can be checked directly, if the dispersions of the flux and charge
have their Gibbsian values (\ref{gi}), the Clausius statement is valid.
This fact has received a special attention in the context of electrical
circuits \cite{landauer,strat}.
More generally, any statistical system which in its stationary state
is described by Gibbs distribution has to satisfy the Clausius formulation
\cite{keizer,klim,balian,AN}.
So it is interesting to ask what will happen with the Clausius formulation
if the temperature
of the bath will be low enough, i.e. in the quantum situation.
Notice that the physical relevance of this question is exactly the same
as in the classical situation, since 
it is expected that thermodynamical relations should not change upon lowering  
the temperature.
As we argued above, the dispersions $\langle \Phi^2\rangle$, 
$\langle Q^2\rangle$ are in general not Gibbsian, 
and the Clausius inequality need not be satisfied. 
Moreover, as was shown in \cite{AN,kapur}
it can be violated in the quantum regime.
Here we will present these results in the context of RLC circuits.

First of all, we notice that there is a general result $\dbarrm \QQ/\d L\ge 0$
valid in all ranges of the parameters \cite{kapur}. 
To see the violation of the Clausius
formulation we  show that one can have
$\d \Sig /\d L\le 0$. We consider low temperatures, i.e.
the quantum frequency $T/\hbar$ is comparable with at least one of other 
frequencies $\omega_0$, $1/(CR)$ and $R/L$ involved in the problem. 
Depending on the value of the quality factor $\omega_0L/R$ 
one can obtain from (\ref{gri1}) two extreme cases \cite{kapur}:
\BEA
&&\frac{\d\Sig ^2}{\d L}=
-\frac{R}{4L^2\omega_0}\ln \left(\frac{\Gamma}{\omega_0}
\right), \quad {\rm for}
\quad \frac{\omega_0L}{R}\gg 1;\\
&&\frac{\d\Sig ^2}{\d L}=-\frac{1}{\pi^2 L}\ln \left(\frac{\Gamma L^2}{CR^3}
\right),\quad {\rm for}
\quad \frac{\omega_0L}{R}\ll 1.
\label{ficus}
\EEA
Recall that $\Gamma$ is assumed to be much larger than any other frequency,
so that both logarithms are positive, implying
that in both cases $\d \Sig /\d L$ is negative.
The first case is realized in case of high quality (weak damping); 
it is then natural that 
$\d \Sig$ is proportional to the small inverse quality, 
since for $R=0$, $\Sigma$ is just
equal to $\half$ (recall that the temperature is low) and, thus, does not vary
with $L$. It is seen also that, apart from a small prefactor, $\d \Sig$ is
multiplied by the logarithm of a large number.
 The second equation describes the
low quality situation, and here $\d \Sigma$ is just proportional to the
logarithm of a large number. This makes the situation especially
interesting, since $L\,\d \Sig^2/\d L$ is at least of order unity. 
For both above cases the change of heat is given by \cite{kapur}:
\BEA
\frac{\dbarrm \QQ}{\d L}=\frac{\hbar R}{2\pi L^2}>0.
\label{kot}
\EEA
Two things have to be noted with this formula: it does not depend on $\Gamma$,
not even through a logarithm, and its ratio to the 
ground-state energy 
$\sim\hbar\omega_0$ of the circuit just produces the quality factor:
$\hbar\omega_0/\Delta \QQ\sim L\omega_0/R$, where $\Delta \QQ\sim L
\dbarrm \QQ/\d L$. So this zero-temperature heat is potentially observable 
for low quality circuits. 
Notice that the very existence of the positive zero-temperature heat 
contradicts the Clausius inequality.

It should be mentioned that there is a widespread 
argument against a positive zero-temperature heat, stating: Since at $T=0$
the bath is in its ground state, it cannot provide energy to the circuit.
This is clearly incorrect, because if the circuit and the bath
do interact, the bath by itself cannot be in its ground state. It is always
in a mixed state, and this is the property of quantum entanglement.
Changing a parameter of the junction can lead to a transfer of
zero-point energy from the bath to the junction, and this should be
identified with heat, since it is arising from the unobservable bath modes.

\renewcommand{\thesection}{\arabic{section}}
\section{Experimental results}
\setcounter{equation}{0}\setcounter{figure}{0} 
\renewcommand{\thesection}{\arabic{section}.}

In the present section we will briefly discuss the possibilities of 
experimental detection of the violation of the Clausius formulation.
In general, one
needs to observe $\langle\Phi^2\rangle$ and $\langle Q^2\rangle$
for several different values of the inductivity $L$. These
are sufficient to recover the corresponding changes of the energy, the
phase-space volume and the work according to formulas (\ref{u}, \ref{sig},
\ref{carbon}) respectively.
In the second step one can check the consistency of the results by
observing directly the work done by the external source, as can be done
using an additional control circuit~\cite{book}. The observed work is 
then subtracted from the total energy to get the heat and
to confirm  $\dbarrm \QQ(T\to 0)\not =0$ and $\dbarrm\QQ/\d L>0$. 
Altogether, the main challenge of the 
experimental observation is in observation of the variances.

We are not aware of experiments which measure 
both $\langle\Phi^2\rangle$ and $\langle Q^2\rangle$ directly. 
However, there are several experiments which
report indirect observations of the variances in different regimes.
In \cite{w1} the authors considered mesoscopic electrical circuits in the 
context of single charge tunneling. The used circuits had thickness of 
the order 10 nm and wideness of the order 1 $\mu$m. The observations 
allowed indirect determination of $\langle Q^2\rangle$.
With the subsequent improving made in \cite{flabby}, 
the correspondence with the
theoretical expression (\ref{gri1}) is perfect. The observations were done
with $C=$4.5 fF, $L=$4.5 nH and for $R$ in the range $10^{1}-10^{3}$ 
k$\Omega$, which corresponds with the quality factor varying from
$10^{-1}$ to $10^{-3}$. 
To avoid thermal noises, the circuits were cooled down to $20$ mK. 
At such a low temperature quantum effects are really dominating, 
since the quantum frequency $T/\hbar\sim 10^{8}$ s$^{-1}$ is comparable 
with the system's characteristic frequencies 
$\omega_0\sim 10^9-10^{10}$ s$^{-1}$, $R/L\sim 10^8$ s$^{-1}$ 
and $1/(RC)\sim 10^{9}$ s$^{-1}$.

Let us now estimate the outcome of our effect with the above parameters.
Taking $R=10^{3}$ k$\Omega$ one gets from (\ref{kot})
$\Delta \QQ\sim L\,\dbarrm \QQ/\d L 
\sim 10^{-19}$ J $\sim 1$ eV, an observable effect. 
On the other hand,  restoring Boltzmann's constant,
the right hand side of the Clausius inequality
$k_BT\Delta S\sim k_BT$  takes a much smaller value,
since for $T=20$ mK one has $k_BT\sim 10^{-25}$J $\sim 10^{-6}$ eV. 
Thus to verify the violation
of the Clausius inequality it suffices to take the sign of
$\Delta L$ positive, which brings a positive $\Delta \QQ$.

\renewcommand{\thesection}{\arabic{section}}
\section{ Conclusion.}
\setcounter{equation}{0}\setcounter{figure}{0} 
\renewcommand{\thesection}{\arabic{section}.}

The present paper discusses the Clausius inequality, one
of the formulations of the second law, in the context of equilibrium
RLC circuits. Following references \cite{AN,kapur} it is confirmed 
that this inequality is broken if the bath temperature is low enough, 
namely, if the characteristic
quantum time-scale $\hbar/T$ is comparable with other relevant times
of the circuit. The result can be briefly summarized as follows: localization
of the system, i.e. decrease of its entropy or phase-space volume, can be
connected with absorption of heat from the bath. This is in a sharp contrast
with the classical experience, where localization occurs with emission of heat.
We provide a simple and sufficiently general formula (\ref{kot}),
which describes the effect at low temperatures. 

One of our main purposes was to compare our result with recent experiments 
done on nanoscale low-temperature circuits \cite{w1,flabby}. 
This comparison led us to conclude
that an experimental verification of the Clausius inequality breaking is
fully within the reach of modern experiments. It is, therefore, hoped that
the present paper will stimulate further experimentation on the issue 
whether non-thermodynamic energy flows occur in nature.

\section*{Acknowledgment}

This work is part of the research programme of the `Stichting 
voor Fundamenteel Onderzoek der Materie', which is financially
supported by the `Nederlandse Organisatie voor Wetenschappelijk
Onderzoek (NWO)'.

\end{document}